\documentclass[pdflatex,sn-mathphys-num]{sn-jnl}

\usepackage{graphicx}%
\usepackage{multirow}%
\usepackage{amsmath,amssymb,amsfonts}%
\usepackage{amsthm}%
\usepackage{mathrsfs}%
\usepackage[title]{appendix}%
\usepackage{xcolor}%
\usepackage{textcomp}%
\usepackage{manyfoot}%
\usepackage{booktabs}%
\usepackage{algorithm}%
\usepackage{algorithmicx}%
\usepackage{algpseudocode}%
\usepackage{listings}%
\usepackage{xspace}
\usepackage{bold-extra}
\usepackage[most]{tcolorbox}
\usepackage{colortbl}

\usepackage{lipsum}
\usepackage{amsfonts}
\usepackage{graphicx}
\usepackage{epstopdf}
\usepackage{amssymb}
\usepackage{bm}
\usepackage{lineno}
\usepackage{textcomp}

\ifpdf
  \DeclareGraphicsExtensions{.eps,.pdf,.png,.jpg}
\else
  \DeclareGraphicsExtensions{.eps}
\fi

\usepackage{xcolor}

\newcommand{\h}{\mathcal{H}}

\usepackage{siunitx}
\DeclareSIUnit{\us}{\micro\second}

\usepackage{textcomp}
\usepackage{microtype}
\usepackage{float}

\begin{document}

\title[Article Title]{A Quantum Annealing Approach for Solving
Optimal Feature Selection and Next Release Problems}


\author[1,3,4]{\fnm{Shuchang} \sur{Wang}}\email{wangsc014@mail.ustc.edu.cn}

\author[2]{\fnm{Xiaopeng} \sur{Qiu}}\email{qiuxiaopeng@mail.ustc.edu.cn}

\author*[1,4]{\fnm{Yingxing} \sur{Xue}}\email{yxxue@ustc.edu.cn}

\author[5]{\fnm{YanFu} \sur{Li}}\email{liyanfu@tsinghua.edu.cn}

\author*[1,3,4]{\fnm{Wei} \sur{Yang}}\email{qubit@ustc.edu.cn}

\affil[1]{\orgdiv{School of Computer Science and Technology}, \orgname{University of Science and Technology of China}, \orgaddress{\city{Hefei}, \postcode{230026}, \state{Anhui}, \country{China}}}

\affil[2]{\orgdiv{School of Software Engineering}, \orgname{University of Science and Technology of China}, \orgaddress{\city{Hefei}, \postcode{230026}, \state{Anhui}, \country{China}}}

\affil[3]{\orgdiv{Suzhou Institute for Advanced Research}, \orgname{University of Science and Technology of China}, \orgaddress{\city{Suzhou}, \postcode{215123}, \state{Jiangsu}, \country{China}}}

\affil[4]{\orgdiv{Hefei National Laboratory}, \orgname{University of Science and Technology of China}, \orgaddress{\city{Hefei}, \postcode{230088}, \state{Anhui}, \country{China}}}

\affil[5]{\orgdiv{Department of Industrial Engineering}, \orgname{Tsinghua University}, \orgaddress{\city{Haidian}, \postcode{100084}, \state{Beijing}, \country{China}}}


\abstract{Search-based software engineering (SBSE) addresses critical optimization challenges in software engineering, including the next release problem (NRP) and feature selection problem (FSP). While traditional heuristic approaches and integer linear programming (ILP) methods have demonstrated efficacy for small to medium-scale problems, their scalability to large-scale instances remains unknown. Here, we introduce quantum annealing (QA) as a subroutine to tackling multi-objective SBSE problems, leveraging the computational potential of quantum systems. We propose two QA-based algorithms tailored to different problem scales. For small-scale problems, we reformulate multi-objective optimization (MOO) as single-objective optimization (SOO) using penalty-based mappings for quantum processing. For large-scale problems, we employ a decomposition strategy guided by maximum energy impact (MEI), integrating QA with a steepest descent method to enhance local search efficiency. Applied to NRP and FSP, our approaches are benchmarked against the heuristic NSGA-II and the ILP-based $\epsilon$-constraint method. Experimental results reveal that while our methods produce fewer non-dominated solutions than $\epsilon$-constraint, they achieve significant reductions in execution time. Moreover, compared to NSGA-II, our methods deliver more non-dominated solutions with superior computational efficiency. These findings underscore the potential of QA in advancing scalable and efficient solutions for SBSE challenges.}




\maketitle

Software engineering is a vast and multifaceted field, encompassing the development, management, and enhancement of software. These considerations frequently involve striking a balance between different objectives that may compete or clash with one another \cite{harman2012search}.
A key challenge is the decision-making process in feature prioritization for software, where cost-effectiveness must be balanced with project requirements. 
Search-based software engineering (SBSE) addresses such challenges by conceptualizing them as search-based optimization problems and employing different kinds of optimal search algorithms derived from operational research to discover feasible solutions.
Among the broad applications of SBSE, this work focuses on two fundamental optimization problems in software requirements engineering and software product line engineering: the Next Release Problem (NRP) and the Feature Selection Problem (FSP), both critical to advancing software engineering practices.

Many optimization problems involve multiple competing objectives that require simultaneous optimization, forming what are known as multi-objective optimization problems. Unlike single-objective optimization, which seeks a single optimal solution, these problems yield a set of solutions, collectively forming the Pareto front \cite{messac2000aggregate}. Within this set, no solution is universally superior to another; improving one objective often comes at the expense of another. The complexity of navigating the Pareto front has driven extensive research, with numerous methodologies proposed to address its inherent challenges. The Pareto front represents an area of interest in multi-objective optimization studies due to its inherent complexity and the challenges it presents in achieving optimal solutions.

NRP and FSP are typical multi-objective optimization problems in software engineering, with numerous schemes proposed for their solution. 
NRP methods primarily fall into two categories: multi-objective evolutionary algorithm (MOEA) based on heuristic thought, and integer linear programming (ILP) based on operations research knowledge.
Early in 2007, NSGA-II was used to solve multi-objective NRP (MONRP) \cite{zhang2007multi}, followed by other evolutionary algorithms, such as genetic algorithms (GA), ant colony optimization (HACO) with local search \cite{jiang2010hybrid}, particle swarm optimization with greedy initialization \cite{hamdy2019greedy}, and a hybrid artificial bee colony and differential evolution method (HABC-DE) \cite{marghny2022hybrid}.
On the ILP side, the $\epsilon$-constraint method was introduced in 2015 to address NRP \cite{veerapen2015integer}, with subsequent improvements like I-EC for bi-objective NRP and SolRep for tri-objective NRP \cite{dong2022multi}.

Similarly, FSP has also been tackled using both MOEA and ILP methods. GA was first leveraged to solve FSP in 2011 \cite{guo2011genetic} and showed good performance. 
The introduction of MOEAs further advanced FSP solutions, with the Improved Binary Evolutionary Algorithm (IBEA) emerging as the most effective MOEA \cite{sayyad2013optimum}. Subsequent hybrid approaches, such as IBED, which combines IBEA with Differential Evolution (DE), further enhanced the performance \cite{xue2016ibed}. Other hybrid strategies included integrating IBEA with constraint solvers, leading to the development of SATIBEA and SMTIBEA \cite{henard2015combining, guo2019smtibea}. In addition to heuristic approaches, ILP methods like $\epsilon$-constraint and SolRep \cite{Xue2018} have been applied to FSP. Notably, researchers have also explored combining ILP with IBEA, resulting in the MILPIBEA method \cite{abdollahzadeh2022multi}.

MOEAs exhibit remarkable versatility, as they can solve problems without necessitating specific domain knowledge.
However, their efficacy diminishes as the problem size increases, due to a sharp rise in computational time \cite{wang2017application}. Moreover, MOEAs often fail to identify all Pareto solutions, limiting their utility. 
In contrast, ILP  holds the potential to find out all Pareto solutions. Nevertheless, it is not suitable for middle- or large-scale combinatorial optimization problems, which are NP-hard problems. As the problem size expands, the time required for finding valid solutions grows exponentially, making ILP less efficient. 
Therefore, when dealing with large-scale problems, both MOEA and ILP grapple with the challenge of finding feasible solutions within a restricted time, hindering their application in complex software engineering environments. This highlights the urgent need for an efficient approach that can deliver high-quality solutions quickly in the context of SBSE.

Quantum annealing (QA) is an emerging optimization technique that forms the focus of this work. The genesis of quantum computing can be traced back to the early 1980s when initial concepts centered on harnessing the power of superposition and entanglement for computational tasks and the simulation of natural phenomena \cite{benioff1980computer}. 
Over several decades, quantum computers have evolved remarkably, with their capacity increasing exponentially from a mere few dozen bits to thousands. This dramatic advancement has led to superior performance across a wide range of applications. Hence, quantum-based algorithms have been observed to provide substantial speed advantages over their classical counterparts \cite{grover1996fast, vodeb2025stirring}.
In the context of combinatorial optimization problems, the application of QA necessitates the formulation of these problems as Quadratic Unconstrained Binary Optimization (QUBO), offering a unified methodology for addressing NRP and FSP. 
Despite the proven efficacy of QA in various fields, its application to SBSE remains largely unexplored. The main challenge lies in effectively leveraging the unique quantum characteristics to develop a feasible and efficient algorithm. Hence, to the best knowledge of ours, this work makes the first endeavors to explore the potential applicability of QA to NRP and FSP.

\begin{figure}[t]
    \centering
    \includegraphics[width=0.8\linewidth]{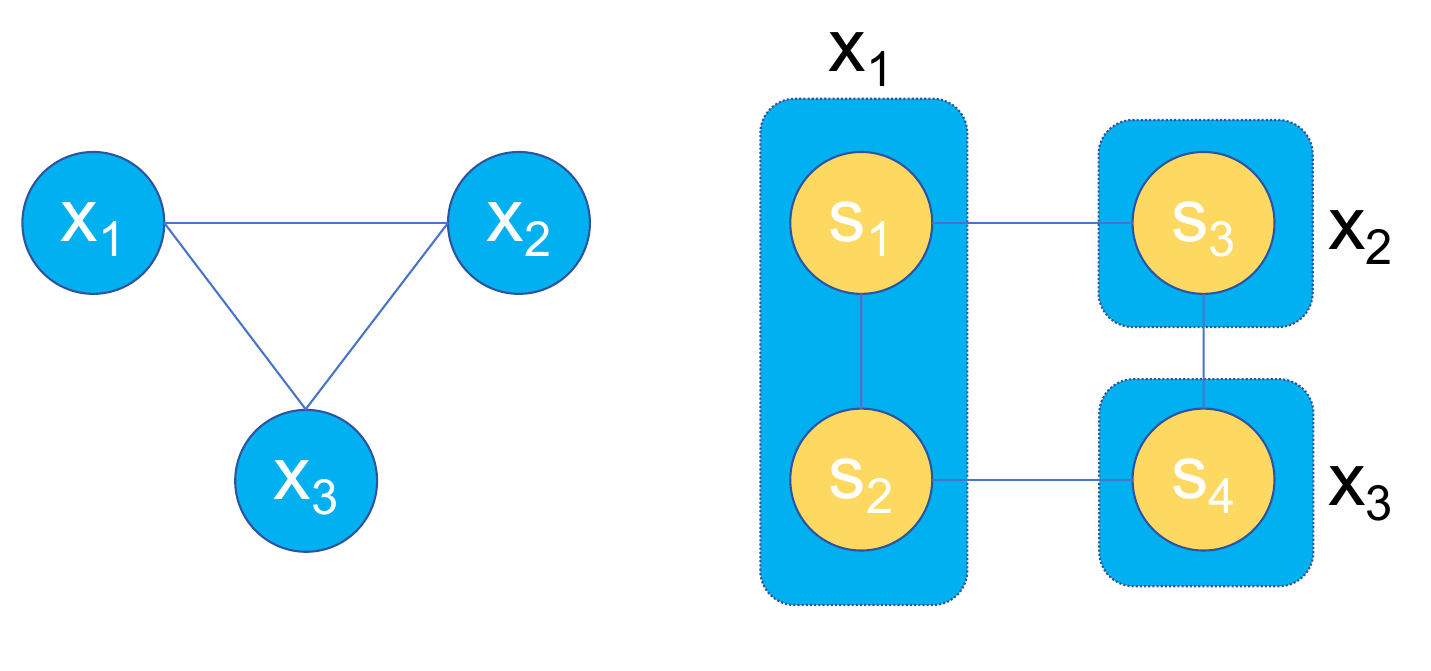}
    \caption{Embedding on QPU. This figure shows a QUBO problem being embedded on QPU.
In the left part of the figure, $\h_{QUBO}$ has linear terms $x_1$, $x_2$, $x_3$ and quadratic terms $x_1x_2$, $x_1x_3$, $x_2x_3$.
And in the right part of the figure, $\h_{QPU}$ is described by qubits $s_1$, $s_2$, $s_3$, $s_4$ and couplers $s_1s_2$, $s_1s_3$, $s_2s_4$, $s_3s_4$.
$x_1$ is mapped to $\{s_1, s_2\}$ and $x_2$, $x_3$ is mapped to $s_3$, $s_4$ separately.
$\{s_1, s_2\}$ is called a \textit{chain} and $s_1$, $s_2$ is restricted with $s_1 = s_2$.
In other words, each variable is mapped to a set of equally evaluated qubits.}
    \label{fig:embed}
\end{figure}

As mentioned before, existing methods for solving SBSE are mainly divided into two categories: MOEA and ILP.
Under such a circumstance, the two solving methods, NSGA-II method and $\epsilon-$constraint method, which are successfully applied to the MOO problem are adopted in this study to provide a fundamental approach to the SBSE problem.

To apply the quantum annealing approach to SBSE, the problem must first be transformed into a QUBO model, and then mapped onto the qubit topology. The utilization of quantum phenomena, such as superposition and entanglement \cite{das2008colloquium}, is instrumental in this process. The ultimate goal is to locate the system energy's lowest state during the annealing procedure. 
This method is primarily effective for resolving single-objective optimization problems. However, multi-objective optimization problems, characterized by multiple conflicting objectives, necessitate a different approach.

In this work, we introduce a novel application of QA to solve SBSE problems, specifically targeting multi-objective NRP and FSP at both small and large scales. To address this, we develop two separate algorithms that capitalize on the capabilities of QA.
For small-scale problems, which fit within the computational capacity of current quantum computers, we propose the Multi-Objective Quantum Annealing (MOQA). MOQA employs a weighted aggregation approach to transform the multi-objective optimization problem into a single objective format, then converts constraints into a QUBO model for QA processing.
On the other hand, for large-scale problems that exceed the direct computational ability of current quantum computers, we introduce the Classical-Quantum Hybrid Annealing (CQHA). CQHA decomposes the large-scale problem into multiple sub-problems using the maximum energy impact (MEI) method \cite{Decompose}, which isolates the influential variables to simplify computation. This method is then combined with the steepest descent method to enhance local search capability. This hybrid approach is expected to yield of a greater number of non-dominated solutions \cite{ozlen2009multi, deng2022enhanced}. In multi-objective optimization, a dominated solution refers to a solution that is inferior to another solution in all objectives. Such solutions are generally considered less desirable, as they are outperformed by other solutions in the search space. The significance of producing dominated solutions lies in their ability to help refine the search process by eliminating less efficient alternatives keeping the non-dominated solutions. By focusing on non-dominated solutions, an algorithm can concentrate on finding the Pareto optimal front, ensuring that the resulting solutions represent the best trade-offs among conflicting objectives.

According to \cite{rahimi2023comparative}, NSGA-II demonstrated the fastest execution time among 19 state-of-the-art MOEA methods. 
To evaluate efficiency,  we compare our QA-based methods against NSGA-II, assessing which methods find  solutions more efficiently without compromising solution quality. 
Additionally, we benchmark against the ILP method $\epsilon$-constraint  in small cases where exact methods perform well. 
The results indicate that, in most cases, QA methods outperform NSGA-II in both small- and large-scale problems, reducing running time by an average of 30.8\% (small cases) and 94.1\% (large cases), respectively. 
For very small cases (fewer than 50 variables), $\epsilon$-constraint is faster than QA, but as problem size grows, its runtime becomes significantly longer—exceeding QA's by more than 100 times.

To sum up, our main contributions are as follows:

\begin{enumerate} 
  \item  
  We convert the nonlinear logic formulae in FSP and NRP into a QUBO model, so that these problems can be solved by a quantum computer. We also give the derivation process of the conversion.
  \item  
   We apply QA to solve multi-objective problems and propose two methods, namely, MOQA for small-scale problems and CQHA for large-scale problems. To our knowledge, the QA-based methods have not been applied to the NRP, FSP, or similar SBSE problems.
  \item  
   We apply the maximum energy impact (MEI) method to decompose the large-scale problem into multiple sub-problems, making them solvable on quantum computers with limited hardware. 
  \item 
   Compared with the state-of-the-art methods (MOEA and ILP), our proposed methods significantly improve the solving efficiency of the resulted non-dominated solutions, supported by the superior experimental results on commonly used NRP and FSP problems. We also utilized the quantum platform MindSpore Quantum in our experiments \cite{xu2024mindspore}. The data and the codes that support the findings of the classical simulations are openly available at the Gitee repository: \url{https://gitee.com/mindspore/mindquantum/tree/research/paper_with_code/a_quantum_annealing_approach_for_solving_NRP_and_FSP}.
\end{enumerate}

\section*{Results}\label{sec2}

\subsection*{Research Questions} 

To gain deeper insight into the performance of our proposed approaches, we seek to address the following research questions (RQs):

\vspace{-1mm}
\begin{enumerate}
\item  On small-scale problems, how would MOQA perform on multi-objective NRP and FSP compared with the ILP method $\epsilon$-constraint and the MOEA method NSGA-II? 
\item  On large-scale problems, how would CQHA perform on multi-objective NRP and FSP compared with NSGA-II? 
\item  How would CQHA behave when discarding some of the lower energy sub-problems?
\item  How would the hybrid algorithm behave when the size of data increases?
\item  How would D-Wave solver behave when different numbers of qubits are available?
\end{enumerate}

\subsection*{Evaluation Setup}

\subsubsection*{Datasets}\label{data}
Our approach to addressing NRP problems utilizes realistic datasets from the ReleasePlanner projects \cite{rp}, specifically the \textit{MSWord} and \textit{RP} datasets.
In these datasets, each stakeholder assigns a \textit{revenue} value to each requirement, while a \textit{cost} is attributed to the corresponding requirements. Revenue computation in our work follows the methodology established in \cite{dong2022multi, revenue}. 
Additionally, we include the \textit{Baan} dataset \cite{Baan} and five synthetic datasets, commonly referred to as the classical datasets \cite{xuan2012solving}(denoted as classic-1 to classic-5), to ensure a comprehensive evaluation. 

Our study addresses FSP by utilizing software product line datasets sourced from Splot Search \cite{splot}, a widely recognized benchmark repository among researchers. It includes multiple datasets such as \textit{WebPortal}, \textit{Drupal}, \textit{BerkeleyDB}, \textit{ERS}, and \textit{E-Shop}. 
Moreover, we incorporate classical datasets from the LVAT project repository \cite{berger2012variability}, specifically \textit{toybox} and \textit{uClinux}. 
For larger datasets like \textit{eCos} and \textit{LinuxX86}, both CQHA and NSGA-II struggle to find feasible solutions, leading to their exclusion from the experiment. 
To strengthen the validity of our experiments, we introduce two artificially generated datasets by SPLOT FM generator \cite{splot}: \textit{SPLOT-1} and \textit{SPLOT-2}. 
Each feature present in these models is characterized by four attributes: \textit{Richness}, \textit{Reliability}, \textit{Defects}, and \textit{Cost} \cite{fourobj}.

The datasets used in our experiments are based on the most classical and widely used benchmarks in NRP and FSP. All selected models are intended to facilitate direct comparisons with prior studies \cite{tan2015optimizing} \cite{xue2016ibed} \cite{Xue2018}\cite{dong2022multi}.

\subsubsection*{Environment}
We utilize Python and D-Wave's Python API to interface with the D-Wave Leap Advantage System 4.1 machine, boasting 5760 active physical qubits.
Furthermore, jMetal in Java is employed for NSGA-II as our MOEA methods, and Cplex(12) \cite{cplex2009v12} is employed as our ILP solver in exact methods. 

Our experiments are conducted on an Ubuntu 18.04 system with an Intel(R) Xeon(R) Platinum 8160 processor and 376 GB of RAM. 

\subsection*{Quality Indicators}
For multi-objective problems, besides execution time and the number of non-dominated solutions for each method, we adopt four additional indicators to assess the quality of the solution set: \textbf{HyperVolume} (HV), \textbf{Inverted Generational Distance} (IGD), \textbf{Spacing} (SP) and \textbf{Num. of Pareto non-dominance} (NoP,  usually given as $\bm{\|N_S\|}$)\cite{multi-objective-indicators, veerapen2015integer, Xue2018}. 
Note that in Tables \ref{tab:MOQA-NSGAII-NRP} through \ref{tab:Hybrid-MOO-Sacle}, $\bm{time(s)}$ denotes the execution time for each method, measured in seconds. $\bm{\|S\|}$ represents the number of correct non-dominated solutions separately found by the corresponding method. $\bm{\|N_S\|}$ denotes the number of correct non-dominated solutions that are not dominated by any method. In other words, it counts the solutions from all methods combined that are Pareto-optimal and not dominated by any solution from any other method. These are the solutions that form the approximate Pareto front.
We first calculate the approximate Pareto front by the union of all methods, and then use it to guide the calculation of $\|N_S\|$, HV, IGD, and SP.


 

\begin{enumerate}
  \item 
  \textbf{\emph{HV:}} HyperVolume quantifies the volume of the region dominated by the solution set. We normalize the objective coefficient before calculating HV, and a higher HV value is preferred. 
  \item 
  \textbf{\emph{IGD:}} Inverted Generational Distance measures how closely the solution set approximates the Pareto front, and a lower IGD value is preferred. 
  \item 
  \textbf{\emph{SP:}} Spacing calculates the standard deviation of the shortest distances between each solution and its nearest solution, and a lower SP value is preferred.
  \item 
  \textbf{\emph{NoP:}} This indicator counts the number of correct non-dominated solutions in the approximate Pareto front, which consists of the solutions set obtained by all methods.
\end{enumerate}

\subsection*{Answer to RQ1}
To answer the question, we implement small-scale problems that could be solved directly by a quantum computer with limited quite support. We compare the results of $\epsilon-$constrain (A), MOQA (B), and NSGA-II (C), respectively. 
$\epsilon-$constrain is an exact method for multi-objective problems which can search the whole Pareto front, and we use it to calculate the number of Pareto solutions of MOQA and NSGA-II, as well as use it as a reference for IGD indicators and HV indicators.

The parameters of NSGA-II follow the work in \cite{veerapen2015integer}, with a population size of 100, crossover probability of 0.8, mutation probability of 1/$n$ (where $n$ is the size of decision variables), evaluation of 20,000 (200 generations). For smaller cases (\textit{RP} and \textit{MSWord}), the evaluation is reduced to 10,000 (100 generations). 
For the MOQA method, we generate the weight $W$ randomly 10 times to formulate problems into different QUBOs, and each QUBO is sampled 500 times an annealing time of 20 $\mu s$, yielding 5,000 times per run. 
For all cases, both NSGA-II and MOQA repeat the run 30 times, and the results are averaged across these runs.
We implement all test cases, and the results are shown in Table \ref{tab:MOQA-NSGAII-NRP} for NRP cases and Table \ref{tab:MOQA-NSGAII-FSP} for FSP cases. In any cases, if the result of a method is not listed in the table, it indicates that the method cannot find a non-dominated solution.

\begin{table}[ht]
    \centering
    \caption{Non-dominated solutions found by $\epsilon-$constrain (A),  MOQA (B) and NSGA-II (C) for small-scale NRP cases. Note that MOQA cannot find a non-dominated solution on \textit{Baan}, and NSGA-II cannot find a non-dominated solution on \textit{classic-1}, so these results are not included here.}
    \vspace{-2mm}
    \label{tab:MOQA-NSGAII-NRP}
    \begin{tabular}{|c|c c c c c c|}
        \hline
        \rowcolor{blue!20} \textbf{RP} & \textbf{time(s)} & \textbf{$\|S\|$}  & \textbf{$\|N_S\|$} & \textbf{IGD} & \textbf{HV} & \textbf{SP}\\ 
        \hline
        \rowcolor{gray!20} A & 0.38 & 71.0 & 71.0 & 0.0 & 0.76 & 313.91\\
        B & 0.64 & 71.0 & 71.0 & 0.0 & 0.76 & 313.91\\
        \rowcolor{gray!20} C & 0.68 & 63.0 & 61.2 & 30.03 & 0.76 & 338.19\\ 
        \hline
        \rowcolor{blue!20} MSWord & \textbf{time(s)} & \textbf{$\|S\|$}  & \textbf{$\|N_S\|$} & \textbf{IGD} & \textbf{HV} & \textbf{SP}\\ 
        \hline
        \rowcolor{gray!20} A & 1.07 & 189.0 & 189.0 & 0.0 & 0.76 & 4.14\\
        B & 0.71 & 143.2 & 110.3 & 2.16 & 0.76 & 4.65\\
        \rowcolor{gray!20} C & 1.23 & 79.2 & 31.6 & 72.99 & 0.17 & 6.99\\ 
        \hline
        \rowcolor{blue!20} Baan & \textbf{time(s)} & \textbf{$\|S\|$}  & \textbf{$\|N_S\|$} & \textbf{IGD} & \textbf{HV} & \textbf{SP}\\
        \hline
        \rowcolor{gray!20} A & 199.46 & 793.0 & 793.0 & 0.0 & 0.54 & 43.21\\
        C & 2.21 & 91.0 & 2.2 & 278.39 & 0.34 & 199.34\\
        \hline
        \rowcolor{blue!20} classic-1 & \textbf{time(s)} & \textbf{$\|S\|$}  & \textbf{$\|N_S\|$} & \textbf{IGD} & \textbf{HV} & \textbf{SP} \\
        \hline
        \rowcolor{gray!20} A & 117.56 & 465.0 & 465.0 & 0.0 & 0.71 & 2.71\\
        B & 0.91 & 225.6 & 55.4 & 8.8 & 0.71 & 6.37 \\
        \hline
    \end{tabular}
\end{table}

From the tables, we can observe that $\epsilon$-constraint always finds all Pareto solutions, achieving an IGD score of 0 and excelling in both HV and SP indicators. MOQA finds all Pareto solutions on \textit{RP} cases, while NSGA-II fails to identify all solutions in any of the test cases. For the smallest case \emph{RP}, $\epsilon$-constraint completes the solution in nearly two-thirds of the time required by MOQA and NSGA-II, though it takes significantly longer for larger datasets, particularly for E-Shop, where it cannot complete within 72 hours.
MOQA generally outperforms NSGA-II in terms of non-dominated solutions, except for the \textit{Baan} case.
On \textit{Baan}, MOQA takes 1.15 \textit{s} but fails to find any non-dominated solution, while NSGA-II finds 2.2 solutions on average in 2.21 s. Compared with $\epsilon$-constraint, both MOQA and NSGA-II face difficulties in solving \textit{Baan}.
MOQA always performs better than NSGA-II on IGD and HV indicators (except for \textit{Baan}), especially on \textit{classic-1}, where NSGA-II takes 5.33 \textit{s} (five times longer than MOQA) but cannot find a non-dominated solution.
In most cases, MOQA performs better than NSGA-II on the SP indicator (except for \textit{ERS}, \textit{Baan} and \textit{E-Shop}), showcasing solutions found by MOQA are more balanced than those found by NSGA-II.
In terms execution times, MOQA reduces the time by 30.8\% on average compared to NSGA-II (excluding Baan) and by 69.5\% compared to $\epsilon$-constraint (excluding E-Shop).


\begin{table}
\vspace{-2mm}
    \centering
    \caption{Non-dominated solutions found by $\epsilon-$constrain (A), MOQA (B) and NSGA-II (C) for small-scale FSP cases. Note that $\epsilon-$constrain cannot complete solving within 72 hours, so these results are not shown in the table.}\label{tab:MOQA-NSGAII-FSP}
    \vspace{-2mm}
    \begin{tabular}{|c|c c c c c c|}
        \hline
        \rowcolor{blue!20} BerkeleyDB & time(s) & $\|S\|$  & $\|N_S\|$ & IGD & HV & SP\\
        \hline
        \rowcolor{gray!20} A & 13.49 & 57.0 & 57.0 & 0.0 & 1.98 & 2.06\\
        B & 0.56 & 54.2 & 54.2 & 0.14 & 1.98 & 2.35 \\
        \rowcolor{gray!20} C & 0.61 & 55.4 & 53.3 & 0.22 & 1.98 & 2.1 \\
        \hline
        \rowcolor{blue!20} ERS & time(s) & $\|S\|$  & $\|N_S\|$ & IGD & HV & SP\\
        \hline
        \rowcolor{gray!20} A & 33.23 & 40.0 & 40.0 & 0.0 & 0.86 & 1.63\\
        B & 0.65 & 33.1 & 17.2 & 1.88 & 0.86 & 2.4 \\
        \rowcolor{gray!20} C & 0.78 & 6.6 & 5.2 & 8.61 & 0.73 & 1.75 \\
        \hline
        \rowcolor{blue!20} WebPortal & time(s) & $\|S\|$  & $\|N_S\|$ & IGD & HV & SP\\
        \hline
        \rowcolor{gray!20} A & 1054.77 & 605.0 & 605.0 & 0.0 & 2.37 & 2.04\\
        B & 0.70 & 241.2 & 113.0 & 9.3 & 2.37 & 2.06\\ 
        \rowcolor{gray!20} C & 1.17 & 69.8 & 6.2 & 17.02 & 0.99 & 3.06 \\
        \hline
        \rowcolor{blue!20} Drupal & time(s) & $\|S\|$  & $\|N_S\|$ & IGD & HV & SP\\
        \hline
        \rowcolor{gray!20} A & 398.45 & 974.0 & 974.0 & 0.0 & 2.6 & 1.39\\
        B & 0.72 & 385.7 & 306.8 & 6.37 & 2.6 & 2.12\\ 
        \rowcolor{gray!20} C & 1.02 & 85.4 & 11.4 & 11.35 & 1.85 & 4.03 \\
        \hline
        \rowcolor{blue!20} E-Shop & time(s) & $\|S\|$  & $\|N_S\|$ & IGD & HV & SP\\
        \hline
        \rowcolor{gray!20} B & 1.12 & 234.4 & 234.4 & 4.59 & 1.84 & 14.14\\ 
        C & 4.70 & 39.0 & 30.8 & 101.1 & 0.35 & 2.8 \\
        \hline
    \end{tabular}
\end{table}


In the context of very small problem sizes, the $\epsilon-$constraint method can efficiently find solutions in a remarkably short time, making execution time less of a limiting factor. Furthermore, the $\epsilon-$constraint method is capable of identifying a broader range of high-quality non-dominant solutions, thereby making it a preferred method for obtaining superior quality solutions. 
However, as the problem scale escalates, the solution time for $\epsilon-$constraint grows exponentially, rendering it challenging to fulfill the solution within a restricted timeframe. This becomes particularly pronounced in the real-world complex process of software engineering, where the time required to find a solution emerges as a critical bottleneck for production efficiency. Under such circumstances, algorithms like NSGA-II and MOQA, which offer accelerated solutions, become more viable options.

Moreover, a comparative analysis between MOQA and NSGA-II reveals that MOQA has a distinct advantage. It not only identifies a greater number of non-dominated solutions in less time but also achieves higher solution quality. Therefore, in terms of both efficiency and solution quality, MOQA proves to be a more effective method than NSGA-II.

Both MOQA and NSGA-II are heuristic algorithms, and in most cases, neither of them can find all Pareto front solutions. However, their solution strategies differ fundamentally. Leveraging quantum computing, MOQA rapidly explores the solution space, enabling faster identification of non-dominated solutions. In addition, MOQA employs multiple sampling iterations with randomly assigned weights, each directing the search along different trajectories. Within each sampling run, both optimal and suboptimal solutions are identified, with the latter having a probabilistic chance of being non-dominated in the original multi-objective problem. This approach allows MOQA to discover a greater number of non-dominated solutions within a shorter timeframe.

Furthermore, compared with NSGA-II, MOQA is globally optimized for each weight, whereas NSGA-II evolves its population iteratively. When NSGA-II converges, the population is concentrated in a certain region of the multi-objective solution space. In contrast, MOQA employs uniformly distributed weights, ensuring a more balanced exploration of the solution space.

The answer to RQ1 can be summarized as follows: For relatively small scale problems, the $\epsilon$-constraint method exhibits superior performance in comparison to both MOQA and NSGA-II by producing a greater number of non-dominated solutions. 
As the complexity and size of the problem escalate, MOQA and NSGA-II outperform the $\epsilon$-constraint method in efficiency, which is critical for software engineering. 
Additionally, when compared to NSGA-II, MOQA 
could find more non-dominated solutions within a shorter time while achieving higher solution quality.

\subsection*{Answer to RQ2}\label{RQ2}
To answer the question, we implement large-scale test cases that cannot be solved directly by a quantum computer with limited qubit support. For these cases, the exact method can hardly complete the solution in a limited time. Therefore, we compare the performance of CQHA with only the NSGA-II algorithm. 

For CQHA, we generate weights randomly 10 times, and each weight is sampled 500 times with 20 $\mu s$ annealing time. 
As noted in Methods section, CQHA's performance varies with different rate parameters. To
facilitate a unified comparison, the $rate$ in CQHA is set to 100\%.
As for NSGA-II, the population size is set to 500, and the evaluation is set to 500,000 (1000 generations). 
We use the solution composition of CQHA and NSGA-II to approximate the Pareto front to calculate metrics such as IGD and HV, which is logical in verifying the solution performance of the two methods. In most NRP cases, it is difficult for NSGA-II to find more than two non-dominated solutions, so the SP indicator is excluded for large-scale NRP cases. The results are displayed in Table \ref{tab:Hybrid-MOO-NRP} for NRP cases, and Table \ref{tab:Hybrid-MOO-FSP} for FSP cases.


\begin{table}
\vspace{-2mm}
    \centering
    \caption{Non-dominated solutions found by CQHA (A) and NSGA-II (B) for large-scale NRP cases.}\label{tab:Hybrid-MOO-NRP}
    \vspace{-2mm}
    \begin{tabular}{|c|c c c c c|}
        \hline
        \rowcolor{blue!20} classic-2 & time(s) &$\|S\|$& $\|N_S\|$ & IGD & HV\\
        \hline
        \rowcolor{gray!20} A & 57.96 & 134.2 & 134.2 & 0.0 & 0.5\\
        B & 2976.17 & 7.6 & 1.3 & 3104.0 & 0.04\\
        \hline
        \rowcolor{blue!20} classic-3 & time(s) &$\|S\|$& $\|N_S\|$ & IGD & HV\\
        \hline
        \rowcolor{gray!20} A & 206.03 & 63.8 & 63.8 & 67.05 & 0.33  \\
        B & 2984.61 & 2.0 & 2.0 & 3360.82 & 0.08  \\
        \hline
        \rowcolor{blue!20} classic-4 & time(s) & $\|S\|$  & $\|N_S\|$ & IGD & HV  \\
        \hline
        \rowcolor{gray!20} A & 599.11 & 40.4 & 40.4 & 105.6 & 0.33 \\
        B & 3113.86 & 1.6 & 1.6 & 3050.84 & 0.12  \\
        \hline
        \rowcolor{blue!20} classic-5 & time(s) & $\|S\|$  & $\|N_S\|$ & IGD & HV \\
        \hline
        \rowcolor{gray!20} A & 296.22 & 84.1 & 84.1 & 16.72 & 0.41 \\
        B & 3474.54 & 3.8 & 2.0 & 4667.36 & 0.1 \\
        \hline
    \end{tabular}
\vspace{-2mm}
\end{table}


\begin{table}
    \centering
    \caption{Non-dominated solutions found by CQHA (A) and NSGA-II (B) for large scale FSP systems.}\label{tab:Hybrid-MOO-FSP}
    \vspace{-2mm}
    \begin{tabular}{|c|c c c c c c|}
        \hline
        \rowcolor{blue!20} toybox & time(s) &$\|S\|$& $\|N_S\|$ & IGD & HV & SP\\
        \hline
        \rowcolor{gray!20} A & 13.60 & 375.0 & 334.8 & 7.17 & 0.1 & 3.82 \\
        B & 1043.66 & 429.8 & 273.4 & 14.39 & 0.08 & 2.75 \\
        \hline
        \rowcolor{blue!20} uClinux & time(s) & $\|S\|$  & $\|N_S\|$ & IGD & HV & SP \\
        \hline
        \rowcolor{gray!20} A & 148.81 & 511.7 & 511.7 & 6.21 & 0.1 & 7.54\\
        B & 3390.72 & 44.0 & 44.0 & 659.28 & 0.1 & 4.08 \\
        \hline
    \end{tabular}
\end{table}

From the tables, it is evident that in most cases, CQHA outperforms NSGA-II by identifying more non-dominated solutions (NSGA-II is difficult to find a solution that is not dominated by CQHA). Especially for \emph{SPLOT-1} and \emph{SPLOT-2} that are not shown in the tables, with the given parameters, NSGA-II takes very long time, but it fails to discover a feasible solution. In contrast, CQHA takes merely 44.58 \emph{s} to find 70.0 non-dominated solutions for \emph{SPLOT-1} and 48.01 \emph{s} to find 29.2 non-dominated solutions for \emph{SPLOT-2}, on average. Compared with NSGA-II, CQHA reduces the running time by 94.1\% while yielding over 30 times more non-dominated solutions.
Moreover, CQHA outperforms NSGA-II in terms of both IGD and HV. The difference in IGD scores between the \emph{classic-5} and \emph{uCLinux} datasets is more than 100 times, indicating that CQHA’s solutions are closer to the Pareto front.

In the FSP cases \textit{toybox} and \textit{uClinux}, NSGA-II performs better on the SP indicator, resulting in lower SP scores, suggesting that it finds more evenly distributed solutions than CQHA.
Decomposing a large-scale problem into multiple subproblems is a lossy decomposition that may lose some global information, potentially leading to over-optimization in certain cases and affecting the performance of CQHA on SP metrics. Moreover, introduction of local search in CQHA makes the solution set of CQHA more concentrated in the solution space. By contrast NSGA-II can balance all objectives globally, which leads to NSGA-II performing better on SP than CQHA.

The answer to RQ2 is as follows: In NRP cases, CQHA identifies more non-dominated solutions in less time, and outperforms NSGA-II on IGD, HV, and SP indicators. As to FSP cases,  CQHA generates more non-dominated solutions in less time, while NSGA-II achieves a more evenly distributed solution set.

\subsection*{Answer to RQ3} \label{RQ3}
To answer the question, when the decomposer breaks down QUBO into multiple subQUBOs, we separately select the top 30\%, 50\%, 70\%, and 90\% of subQUBOs based on their MEIs to participate in the QA process, thereby exploring the performance of CQHA under different $rate$ parameters. 
The values of the remaining variables that do not participate in the QA process are solved using ClassicalSolver with initial values. 
In this experiment, we select the test cases \emph{classic-3}, \emph{classic-4}, \emph{toybox} and \emph{uClinux}, and the results are shown in Table \ref{tab:Hybrid-MOO-Sacle}.



\begin{table}[ht]
\vspace{-2mm}
    \centering
    \caption{Measure CQHA with different sub-QUBOs $\bm{rate}$ in multi-objective problems.}\label{tab:Hybrid-MOO-Sacle}
    \vspace{-2mm}
    \begin{tabular}{|c|c c c c c c|}
        \hline
        \rowcolor{blue!20} classic-2 & time(s) &$\|S\|$& $\|N_S\|$ & IGD & HV & SP\\
        \hline
        \rowcolor{gray!20} 30\% & 25.44 & 128.3 & 30.1 & 140.83 & 0.49 & 45.94\\
        50\% & 42.33 & 128.7 & 5.4 & 114.98 & 0.53 & 49.78\\
        \rowcolor{gray!20} 70\% & 51.46 & 99.6 & 11.1 & 171.53 & 0.55 & 45.98\\
        90\% & 66.54 & 156.3 & 145.5 & 223.09 & 0.58 & 25.85\\
        \rowcolor{gray!20} 100\% & 57.96 & 134.2 & 9.7 & 142.37 & 0.54 & 37.12\\
        \hline
        \rowcolor{blue!20} classic-3 & time(s) &$\|S\|$& $\|N_S\|$ & IGD & HV & SP\\
        \hline
        \rowcolor{gray!20} 30\% & 72.75 & 85.3 & 35.1 & 32.05 & 0.28 & 25.54\\
        50\% & 119.96 & 78.5 & 26.8 & 95.73 & 0.26 & 26.81\\
        \rowcolor{gray!20} 70\% & 127.78 & 80.2 & 21.4 & 100.51 & 0.27 & 36.38\\
        90\% & 178.25 & 70.6 & 18.3 & 170.48 & 0.28 & 23.09\\
        \rowcolor{gray!20} 100\% & 206.03 & 63.8 & 20.4 & 181.33 & 0.26 & 20.42\\
        \hline
        \rowcolor{blue!20} toybox & time(s) &$\|S\|$& $\|N_S\|$ & IGD & HV & SP\\
        \hline
        \rowcolor{gray!20} 30\% & 4.74 & 617.5 & 286.2 & 13.6 & 0.17 & 4.03\\
        50\% & 7.79 & 892.8 & 603.6 & 21.68 & 0.17 & 13.07\\ 
        \rowcolor{gray!20} 70\% & 8.06 & 672.6 & 534.5 & 73.45 & 0.17 & 11.44\\
        90\% & 11.66 & 593.3 & 508.4 & 24.37 & 0.17 & 12.68 \\
        \rowcolor{gray!20} 100\% & 13.60 & 375.0 & 182.8 & 90.41 & 0.12 & 10.9\\
        \hline
        \rowcolor{blue!20} uClinux & time(s) &$\|S\|$& $\|N_S\|$ & IGD & HV & SP\\
        \hline
        \rowcolor{gray!20} 30\% & 54.17 & 1114.0 & 1017.2 & 24.09 & 0.15 & 8.36\\
        50\% & 90.12 & 554.6 & 397.4 & 276.31 & 0.14 & 9.39\\ 
        \rowcolor{gray!20} 70\% & 107.51 & 673.3 & 185.6 & 145.87 & 0.12 & 8.12\\
        90\% & 146.01 & 493.4 & 51.8 & 245.87 & 0.11 & 9.19 \\
        \rowcolor{gray!20} 100\% & 148.87 & 511.7 & 113.2 & 313.71 & 0.09 & 7.54\\
        \hline
    \end{tabular}
\end{table}

The percentage in the first column of Table \ref{tab:Hybrid-MOO-Sacle} indicates the $rate$ of sub-QUBOs participating in the QA process.
From Table \ref{tab:Hybrid-MOO-Sacle}, we can observe that CQHA may find more non-dominated solutions (given as $\|N_S\|$) in a shorter time when some sub-QUBOs are discarded. \textbf{However, the performance of CQHA varies with different $\bm{rates}$, and there is no clear pattern, so we set the $\bm{rate}$ to 100\% in the Methods section}.
In \textit{classic-2}, using 90\% sub-QUBOs to participate in the QA process gets the largest number of non-dominated solutions, and performs best on IGD and SP indicators. However, it also takes the longest time, even exceeding 100\% of the sub-QUBOs. Besides, 90\% rate performs worst on the HV indicator, suggesting that the steepest descent method requires more time for local search. In addition, setting the $rate$ to 30\% or 70\% can find more non-dominated solutions in a shorter time than the complete CQHA.
In \textit{classic-3} and \textit{uClinux}, only when the $rate$ is set to 90\% does the performance lag behind the complete CQHA. In other cases, a faster search for more non-dominated solutions is achieved, supported by the indicators IGD and HV. Nevertheless the performance is opposite on the SP indicator, with the complete CQHA achieving best scores.
In \textit{toybox}, when the $rate$ is set below 100\%, more non-dominated solutions can be found in a less time, and both IGD and HV indicators show improvement. However, when the $rate$ is set to 50\%, 70\%, and 90\%, its performance on SP is worse than the 100\% case. 

The outcome aligns with practical experience. 
In CQHA, we decompose a large-scale QUBO into smaller sub-QUBOs by the variable's maximal energy impact to enable the quantum computer to solve them. Besides, this decomposition also destroys the problem's structure, which will affect the problem-solving. The influence of each sub-QUBO on the result varies. Discarding some sub-QUBOs with minimal impact can speed up CQHA without reducing the quality of the solution. 
However, discarding some sub-QUBOs will also narrow the algorithm’s focus, which may affect the uniformity of the solution distribution. 

To summarize, the answer to RQ3 is that discarding some sub-QUBOs with a lower energy impact can accelerate CQHA and find more non-dominated solutions, but the distribution of these solutions may be less uniform in the target space.

\subsection*{Answer to RQ4} \label{RQ4}

To answer RQ4, we conduct experiments using the hybrid solver on test cases \emph{uClinux} and \emph{classic-4}, with varying data sizes. They represent NRP and FSP problems separately as we used in the former RQs. The goal is to understand how the performance of the hybrid CQHA is affected by increasing data sizes. As expected, the decomposer proves time-consuming for small-scale problems, where further size reduction is unnecessary. The elapsed times for different data sizes are presented in Table \ref{tab:Performance of sizes}. 

\begin{table}[h!]
\vspace{-2mm}
    \centering
    \caption{Performance of CQHA with increasing data sizes.
    }\label{tab:Performance of sizes}
    \vspace{-2mm}
    \begin{tabular}{|c|c c|}
    \hline
    \rowcolor{blue!20} \textbf{Problem} & \textbf{Size} & \textbf{Elapsed Time (s)} \\
    \hline
    \rowcolor{gray!20} uClinux & 100 & 33.37 \\
    uClinux & 500 & 8.63 \\
    \rowcolor{gray!20} uClinux & 700 & 6.46 \\
    uClinux & 1000 & 4.73 \\
    \rowcolor{gray!20} uClinux & 1200 & 4.69 \\
    \hline
    classic-4 & 100 & 268.56 \\
    \rowcolor{gray!20} classic-4 & 500 & 53.96 \\
    classic-4 & 700 & 41.62 \\
    \rowcolor{gray!20} classic-4 & 1000 & 49.66 \\
    classic-4 & 1200 & 41.86 \\
    \hline
    \end{tabular}
\end{table}

For \emph{uClinux}, the elapsed time decreases as the problem size increases up to 1000, indicating that it becomes more efficient with larger sizes. Moreover, the elapsed time stabilizes around 4.69 seconds at size 1200, indicating a performance plateau. Similarly, for \emph{classic-4}, the elapsed time decreases significantly from size 100 to 500, , but increases slightly at size 1000 and then stabilizes at size 1200. This suggests that while the hybrid CQHA handles larger problem sizes more efficiently initially, it may encounter challenges that slightly reduce its efficiency at higher sizes, particularly for problems with more constraints and complex variable interactions like \emph{classic-4}. As a conclusion, the performance improves with increasing problem sizes up to a certain point, after which the efficiency stabilizes or slightly decreases. 


\subsection*{Answer to RQ5}\label{RQ5}

To address this question, we need to explore the solving capability of the D-Wave solver, by testing QUBO problems with random initialization of different sizes. Gaining deeper insight into the computational capacity of the D-Wave quantum computer enables more efficient solution strategies when scaling data and algorithms. The results are presented in Table \ref{tab:dwave_performance}. 

\begin{table}[h!]
\vspace{-2mm}
\centering
\caption{Performance of D-Wave solver on QUBO Pproblems with different number of qubits.}
\label{tab:dwave_performance}
\vspace{-2mm}
\begin{tabular}{|c|c|c|c|}
\hline
\rowcolor{blue!20} \textbf{Problem Size} & \textbf{QPU Sampling} & \textbf{QPU Access} & \textbf{QPU Programming} \\
\rowcolor{blue!20} \textbf{(qubits)} & \textbf{Time ($\mu$s)} & \textbf{Time ($\mu$s)} & \textbf{Time ($\mu$s)} \\
\hline
\rowcolor{gray!20} 50  & 12984.0 & 28744.37 & 15760.37 \\
100 & 20296.0 & 36057.17 & 15761.17 \\
\rowcolor{gray!20} 150 & 25686.0 & 41449.17 & 15763.17 \\
200 & {Failed to solve} & - & - \\
\hline
\end{tabular}
\end{table}

The results provide key insights into the computational capacity of the D-Wave solver. It successfully solves problems up to 150 qubits but fails at 200 qubits due to embedding issues, highlighting the increasing challenge of qubit connectivity as problem size grows. 

Besides that, we analyze various time metrics to further assess performance:

\begin{itemize}
    \item \textbf{QPU Sampling Time}: This metric quantifies the time required by the quantum processing unit (QPU) to perform the QA process for a given QUBO problem. As the problem size increases, the QPU sampling time rises correspondingly. This behavior is expected because larger problems involve more qubits and interactions, requiring more time for the QA process to explore the solution space and find the optimal or near-optimal solutions. For instance, the QPU sampling time for a 50-qubit problem is approximately 12984.0 $\mu$s, while for a 150-qubit problem, it doubles to 25686.0 $\mu$s. This trend highlights the direct correlation between problem size and the time required for quantum sampling.

    \item \textbf{QPU Access Time}: This metric captures the total time taken for the QPU to become available for a new computation, including overhead from accessing the quantum processor. Similar to the QPU sampling time, the QPU access time increases with the problem size, indicating that larger problems impose greater demands on the quantum hardware. Besides, a consistent overhead is observed in the QPU access time across different problem sizes. For example, the QPU access time for a 50-qubit problem is 28744.37 $\mu$s, while for a 150-qubit problem, it rises to 41449.17 $\mu$s. This suggests that while the access time is partly dependent on the problem size, fixed overheads related to system operation and communication also contribute significantly.

    \item \textbf{QPU Programming Time}: This metric measures the time required to program the QPU with a specific problem instance, including embedding the QUBO problem into the hardware's qubit connectivity graph. Interestingly, the QPU programming time remains relatively constant across different problem sizes, with values around 15760 $\mu$s. This suggests that the programming phase, which includes setting up the problem on the QPU and configuring the necessary parameters, is not significantly affected by the problem size. Once the embedding process is completed, the time required to program the QPU does not scale with the number of qubits involved in the problem.

\end{itemize}

When the problem size increases to 200 qubits, the solver encounters difficulties. The inability to solve the 200-qubit problem is primarily due to challenges to find a suitable embedding. This is likely caused by the limited connectivity between qubits in the D-Wave machine, making it difficult to embed larger problems with dense connectivity. In conclusion, the D-Wave solver is effective for solving QUBO problems up to 150 qubits, and embedding challenges prevent successful problem solving beyond this size. Improvements in problem sparsity and qubit connectivity may help overcome these challenges, allowing for the solving of larger QUBO problems.

\vspace{-2mm}

\section*{Discussion}\label{sec3}

\textbf{Threats to Validity.} 
The datasets employed in the preceding analysis predominantly derive from early-stage projects, which may not fully reflect the complexity and scale of contemporary software systems. Consequently, it becomes essential to procure up-to-date, real-world datasets and develop representative models for more accurate and rigorous evaluation.

Another potential threat lies in the parameters of NSGA-II and QA. Comprehensive enumeration of all parameters to identify the optimal solution effect remains elusive, despite in this study, multiple experimental iterations are undertaken to mitigate errors induced by stochasticity. The plethora of QA parameters, coupled with the absence of a foundational guiding principle, renders the examination of individual parameter adjustment impacts on the problem somewhat deficient. It remains conceivable that an ideal parameter configuration exists for a specific dataset, potentially enhancing the efficacy of these algorithms for particular challenges.


\noindent
\textbf{Generality of Quantum Annealing.} 
The inherent constraints of QA hardware present significant challenges for executing large-scale annealing operations. Within the CQHA framework, decomposing large-scale problems into sub-problems follows a lossy decomposition methodology, with the size of these sub-problems being inherently capped. As the magnitude of the primary challenge escalates, the quantity of derived sub-problems concomitantly expands, leading to a pronounced divergence from the original problem and augmenting the complexity of procuring a viable solution.

With the advancement of quantum computing, it is anticipated that quantum annealing technology will witness significant hardware scalability and improved QA scheduling capability. This will enable the solution of larger-scale problems and the design of more complex solution processes. 
Consequently, multi-objective quantum algorithms will benefit from more flexible weight selection strategies. Furthermore, when the quantum annealer possesses stronger initialization capabilities, it would be possible to design hybrid algorithms that interact with classical computers, resulting in solution efficiency that surpasses what can be achieved by classical computers alone.

\section*{Methods}\label{sec4}

\begin{figure}[h!]
    \centering
    \includegraphics[width=0.7\linewidth]{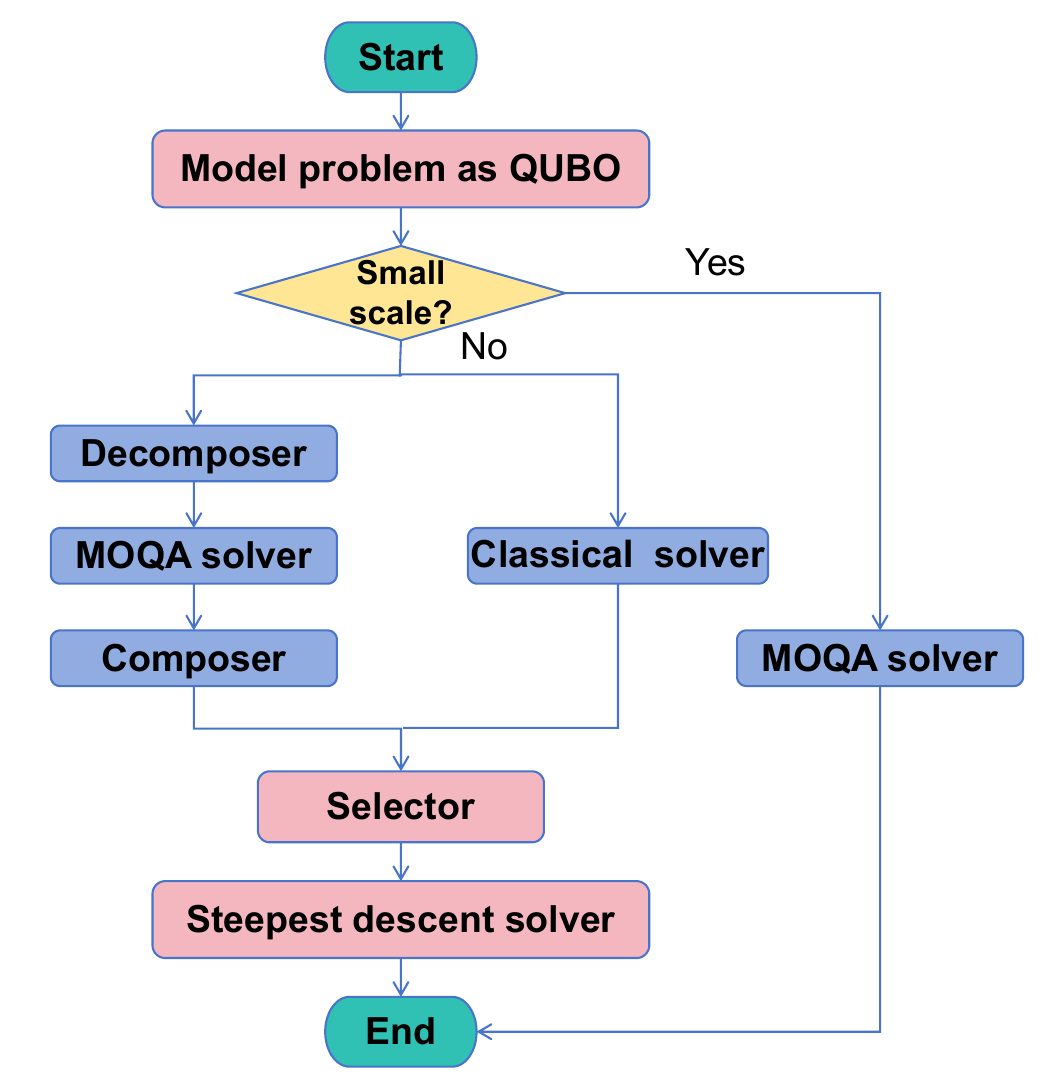}
    \caption{Process of Quantum Annealing.  For small-scale problems, we propose MOQA as the solution method. The significance of this study lies in acknowledging the inherent limitation of present-day quantum computing technology, whereby large-scale problems cannot be feasibly accommodated for quantum computation directly. For large-scale problems, we propose CQHA to apply QA effectively.}
    \label{fig:process}
\end{figure}

The overall process of using QA to solve the minimization problem is outlined in Figure \ref{fig:process}.

\subsection*{MOQA}
\label{moqa}
MOQA involves random weights for objectives aggregation to model different QUBOs, followed by multiple sampling iterations for each QUBO using QA. After repetitive sampling, the quantum computer generates a set of solutions, from which non-dominated solutions are selected. 

The algorithm follows the steps outlined in Algorithm \ref{MOQA}.
Note that \textit{scale($F_i$)} in line 4 scales the objectives by normalization, and \textit{randomWeight()} in line 7 generates a random weight. In line 8 \textit{ModeltoQUBO($F^{\prime}, W, G$)} is responsible for converting the original problem into a QUBO model with currently generated weight $W$. In line 9 $EmbedtoQuantumComputer(QUBO)$ maps QUBO to a specific qubit, and \textit{sample(bqm, r)} in line 10 resamples with QA to get $r$ solutions.
The algorithm obtains the solution set $S^\prime$ and incorporates it into the total solution set $S$. Finally, the algorithm performs the non-dominated sort algorithm $nonDominatedSetSort(F, S)$ in line 13 to obtain the non-dominated solution set as the result.

To achieve a more evenly distributed solution, MOQA generates a series of random weights ($n$ times) and finds multiple multi-objective optimal solutions ($r$ times) for each weight. 
The purpose is not only to identify all optimal solutions within the weight, but also to uncover suboptimal solutions that possibly serve as the multi-objective optimal solutions for the original problem.
MOQA thus sample $n \times r$ samples, with a time complexity of $O(nr)$ for the sampling phase and $O((nr)^2)$ in the non-dominated sorting process. Overall, the inter-complexity of the algorithm is $O((nr)^2))$.

\begin{algorithm}[t]
\caption{MOQA for small-scale instances}
\label{MOQA}
\begin{algorithmic}[1]
    \State \textbf{Input:} $F$: objectives, $G$: constraints, $n$: the number of weights, $r$: sample times
    \State \textbf{Output:} $S$: solutions

    \For{$i \leftarrow 1$; $i \leq M$; $i \leftarrow i + 1$}
        \State $F_{i}^{\prime} \leftarrow$ scale($F_i$)
    \EndFor
    
    \For{$k \leftarrow 1$; $k \leq n$; $k \leftarrow k + 1$}
        \State $W \leftarrow$ randomWeight()
        \State $QUBO \leftarrow$ ModeltoQUBO($F^{\prime}, W, G$)
        \State $bqm \leftarrow$ EmbedtoQuantumComputer($QUBO$)
        \State $S^{'} \leftarrow$ sample($bqm$, $r$)
        \State $S \leftarrow S \cup S^{\prime}$
    \EndFor
    
    \State $S \leftarrow$ nonDominatedSetSort($F$,$S$)\\
    \Return $S$
\end{algorithmic}
\end{algorithm}

\subsection*{CQHA}
\vspace{1mm}

\begin{algorithm}[t]
\caption{CQHA for large-scale instances}
\label{CQHA}
\begin{algorithmic}[1]
    \State \textbf{Input:} $F$: scaled objectives, $G$: constraints, $n$: the number of weights, $r$: sample times, $s$: maximum number of variables for sub-QUBO
    \State \textbf{Output:} $S$: solutions

    \For{$k \leftarrow 1$ to $n$}
        \State $W \leftarrow$ RandomWeight()
        \State $S^{\prime} \leftarrow$ RandomInitialSolution()
        \State $QUBO \leftarrow$ ModeltoQUBO($F, W, G$)
        \State $subQUBOs \leftarrow$ Decomposer($QUBO, rate, s$)
        \While{loop $t$ times or algorithm is not converged}
            \For{qubo in $subQUBOs$}
                \State $S_{sub} \leftarrow$ QuantumSolver($qubo$)
                \State $S_{new} \leftarrow$ Composer($S^{\prime}, S_{sub}$)
                \If{$S_{new}$ gets lower energy than $S^{\prime}$}
                    \State $S^{\prime} \leftarrow $ $S_{new}$
                \EndIf
            \EndFor
            \State $S^{\prime} \leftarrow$ ClassicalSolver($S^{\prime}$)
        \EndWhile
        \State $S \leftarrow S \cup S^{\prime}$
    \EndFor
    \State $S \leftarrow$ nonDominatedSetSort($F, S$)\\
    \Return $S$
\end{algorithmic}
\end{algorithm}

Given current hardware limitations \cite{tran2016hybrid, abbas2024challenges}, quantum computers cannot solve large-scale problems directly \cite{HyOv}. To address this, we introduce CQHA, which decomposes large-scale problems into sub-problems based on the MEI method and employs QA to solve them. Algorithm \ref{CQHA} outlines the general steps of CQHA to solve large-scale problems.

In Algorithm \ref{CQHA},
$RandomWeight()$ generates weight $W$ randomly (line 4), where
$RandomSolution()$ produces a randomly generated initial solution $S^{\prime}$ (line 5). $S^{\prime}$ is a global variable, updated whenever a newly obtained solution dominates it.
With a given weight $W$, scaled objective $F$, and constraints $G$, $ModeltoQUBO(F, W, G)$ convert the problem into a QUBO model (line 6).
In line 7, $Decomposer(QUBO, rate, s)$ applies the MEI method to decompose the QUBO model into multiple sub-QUBOs. 
To enhance efficiency , only sub-QUBOs with significant energy impact are selected for the QA process. The parameter $rate$ controls the proportion of sub-QUBOs participated in the QA process, with 100\% indicating that all sub-QUBOs generated by the decomposer are sent to QuantumSolver. The parameter $s$ indicates the maximum number of variables in each sub-QUBO.
The maximum energy impact method will be explained in detail with an example in section \ref{decomposer}. 
Each sub-QUBO obtained by the decomposer will be solved separately using QA, and the obtained solution $S_{sub}$ is used to replace the initial solution $S^{\prime}$ to obtain a new solution $S_{new}$.
In line 10, $QuantumSolver(qubo)$ uses the QA algorithm to solve each sub-QUBO.
$Composer(S^{\prime}, S_{new})$ then uses the solution of sub-QUBO $S_{sub}$ to replace the initial solution $S^{\prime}$ (line 11), and obtains a new solution $S_{new}$. 
If $S_{new}$ dominates $S^{\prime}$, CQHA updates accordingly (lines 12-14).
To enhance local search capabilities, $ClassicalSolver(S^{\prime})$ uses a classically local solver to further search the solution space of $S^{\prime}$ in line 16 (this work selects the steepest descent algorithm as the local solver).
The process iterates until either a predefined number of iterations $t$ is reached or convergence is achieved. The solution $S^{\prime}$ obtained in each iteration will be merged into the solution set $S$ (line 18). 
Finally, $nonDominatedSetSort(F, S)$ select the non-dominated solution set from $S$ as the result.


\begin{algorithm}[t]
\caption{Process of Decomposer}
\label{decomposeprocess}
\begin{algorithmic}[1]
    \State \textbf{Input:} Q: QUBO, rate: ratio of sub-QUBO participating in QA process, s: the sub-QUBO's maximum variable size.
    \State \textbf{Output:} $SQ$: sub-QUBOs set
    $SQ \leftarrow \varnothing$
    \While{Q remaining variables are not selected}
        \State $p \leftarrow SelectMaxEnergyVar(Q)$
        \State $SQ^{\prime} \leftarrow p$
        \While{Size($SQ^{\prime}$) $<$ s}
            \State $v \leftarrow SelectConnectedVar(p)$
            \State $p^{\prime} \leftarrow SelectMaxEnergyVar(v)$\;
            \\// $pp^{\prime}$ represents the quadratic term of multiplying $p^{\prime}$ by $p$
            \State $SQ^{\prime} \leftarrow SQ^{\prime} + p^{\prime} + {pp^{\prime}}$
            \State $p \leftarrow p^{\prime}$
        \EndWhile 
        \State $SQ \leftarrow SQ \cup SQ^{\prime}$
    \EndWhile
    \State $num \leftarrow Size(SQ) * rate$\;
    \State $SQ \leftarrow SelectMaxEnergySQ(SQ, num)$\;\\
    \Return $S$
\end{algorithmic}
\end{algorithm}
 \vspace{-2mm}

\subsubsection*{Decomposer}
\label{decomposer}

Algorithm \ref{decomposeprocess} provides a detailed explanation of $Decomposer(QUBO, rate, s)$ in line 7 of CQHA.
The decomposition process aims to partition a QUBO into multiple sub-QUBOs based on maximum energy impact, ensuring the resulting sub-problems remain computationally feasible for quantum annealing. The process begins with $SelectMaxEnergyVar(Q)$, which first selects a variable $p$ with the largest coefficient from variables that have not been selected in QUBO $Q$ in line 4, and then incorporates $p$ into sub-QUBO $SQ^{\prime}$ in line 5. 
Loop the lines 7-11 until the variable size of a sub-QUBO reaches the set value $s$.
In line 7, $SelectConnectedVar(p)$ identifies all the variables $v$ along the shared edge (quadratic terms) connected to the selected variable $p$.
In line 8, $SelectMaxEnergyVar(v)$ selects the variable $p^{\prime}$ with the largest coefficient from the variable set $v$. The selected variable $p^{\prime}$ and the quadratic term, $pp^{\prime}$, are incorporated into sub-QUBO $SQ^{\prime}$ in line 10.
Lines 3 to 14 complete the decomposition of a sub-QUBO $SQ^{\prime}$. We loop through these steps until all variables in $Q$ are selected, and multiple sub-QUBOs are obtained and stored in $SQ$.
We calculate the number of sub-QUBOs participating in the QA process $num$ based on $rate$ (line 15), and select the $num$ sub-QUBOs with the highest energy to finally participate in the QA process (line 16).



It is important to note that the MEI decomposition is inherently lossy. To maintain as much of the original problem structure as possible, we set the initial sub-QUBO size $s$ to be as large as possible. If the sub-QUBO cannot be embedded into the quantum hardware, the size is progressively reduced in small increments until a valid embedding is achieved. 

For illustration of how Decomposer works, consider the QUBO which consists of 12 variables. Our goal is to decompose this $QUBO$ into 2 sub-QUBOs, with the maximum variable size for each subQUBO being 6. 
To achieve this, we begin by identifying the variable $x_{i}$ with the largest energy impact. Next, we traverse the unselected variables connected to $x_{i}$ and select five variables with the highest energy values to form $subQUBO_1$. It follows that the remaining variables from a separate sub-QUBO, named $subQUBO_2$. Finally, we send these two sub-QUBOs to the quantum computer for solving, and then construct the final solution to the initial problem using Composer.

In summary, the decomposition strategy balances problem complexity and hardware constraints, enabling QA to handle larger-scale optimization tasks effectively.

\section*{Declarations}

\subsection*{Funding}

This work is supported by the Innovation Program for Quantum Science and Technology (Grant No. 2021ZD0302900).

\subsection*{Data availability}

All datasets are available from the ref. shown in Section $Results$.

\subsection*{Code availability}

The QA-based solver and experimental raw data (This part is based on D-wave API) are on the GitHub repository: \url{https://anonymous.4open.science/r/code_for_paper}. The data and the codes that support the findings of the classical simulations are openly available at the Gitee repository: \url{https://gitee.com/mindspore/mindquantum/tree/research/paper_with_code/a_quantum_annealing_approach_for_solving_NRP_and_FSP}.

\subsection*{Author contribution} 

 S.W. and X.Q. contributed equally to this work. S.W. designed and conducted the experiments, analyzed the results, and drafted the manuscript. X.Q. developed the theoretical framework and participated in the implementation of algorithms. Y.X. contributed to the experiment design, data analysis, and project supervision. Y.L. provided expert insights on industrial engineering perspectives and reviewed the manuscript. W.Y. conceived the project, provided overall guidance, and revised the manuscript critically for intellectual content. All authors discussed the results and approved the final manuscript.

\subsection*{Competing interests}

The authors declare no competing interests.

\end{document}